\documentclass[aps,prb,twocolumn,groupedaddress,showpacs]{revtex4}
\usepackage{graphicx}
\usepackage{amsmath,amsfonts}

\begin{document}

%\preprint{}

\title{Fermi surface and electron dispersions of PbTe doped with resonant Tl impurity from KKR-CPA calculations.}


\author{Bartlomiej Wiendlocha}
\email[e-mail: ]{wiendlocha@fis.agh.edu.pl}
%\thanks{}
\affiliation{Faculty of Physics and Applied Computer Science, AGH University of Science and Technology, Al. Mickiewicza 30, 30-059 Krakow, Poland}


\date{\today}

\begin{abstract}
We present results of detailed study on the electron dispersions and Fermi surface of lead telluride doped with 2\% of thallium, which is resonant impurity in PbTe. Using the KKR--CPA method, Bloch spectral functions (BSFs), which replace the dispersion relations in alloys, are calculated, and BSFs intensity maps over the Brillouin zone (alloy Fermi surface cross-sections) are presented. It is shown, that close to the valence band edge, Tl does not create an isolated impurity band, but due to its resonant character, strongly disturbs the host electronic bands, leading to disappearance of sharp and well defined electronic energy bands.  Consequences of this effect on the transport properties are discussed and new qualitative explanation of the improvement in thermoelectric properties of PbTe:Tl is suggested.

\end{abstract}


\maketitle



%%%%%%%%%%%%%%%%%%%%%%%%%%%%%%%%

\section{Introduction}
Pure and doped PbTe is one of the most studied narrow gap semiconductors\cite{ravich-book} due to its very interesting physical properties and excellent thermoelectric (TE) performance in the mid-temperature range. For both {\it n} and {\it p}-type materials, in the temperature range of 500 -- 800 K it's thermoelectric figure of merit, $zT = \sigma S^2/\kappa$ reaches\cite{ees-jaworski,heremans-science,snyder-n-pbte-ees,snyder-p-pbte-ees} values greater than 1.5 ($\sigma$ is the electrical conductivity, $S$ the thermopower and $\kappa$ the thermal conductivity).
Although PbTe has been studied for many years, still it gains much attention - recently in the context of topological insulators \cite{bansil_snte_topol,pbte_hetero_topol,pressure_topol_pbte} or due to giant anharmonicity.\cite{pbte-anharmonic-nature}
As far as the doped PbTe is concerned, the most interesting physical properties are observed for the group-III impurities,\cite{ravich-review,ravich-review2} among which Tl has a special place. 
At low temperatures, below $T_c \sim 1.5$~K (with $T_c$ depending on Tl concentration) thallium doped lead telluride becomes a BCS--like type II superconductor,\cite{matsushita-supercond} however, the charge Kondo effect and negative-{\it U} Anderson model have been proposed as an explanation for the occurrence of superconductivity,\cite{dzero-prl,matsushita_surf_impedance} and also for the resistivity and thermopower behaviors at low temperatures.\cite{matsushita-kondo,zlatic-prl2012,matsushita_res_kondo}
This Kondo--type behavior is usually connected to the hypothesis of mixed valence state of Tl in PbTe, where Tl is expected to be in Tl$^+$ and Tl$^{3+}$ states. However, in recent XANES and EXAFS measurements\cite{pbte-tl-xanes,pbte-tl-exafs} Tl was found to be only in Tl$^+$ state, thus this issue needs further investigation. At the same time, it has been known for the long time that Tl act as a resonant impurity in PbTe.\cite{ravich-review,ravich-review2,ees-review}
Existence of Tl in Tl$^+$ and the formation of resonant level (RL) are supported by the {\it ab initio} DFT calculations for PbTe:Tl,\cite{ees-jaworski,mahanti-prl06,mahanti-prb08,ees-review,takagiwa-tlpbte} where the formation of the two resonant levels, associated with two Tl 6s electrons, was found.
The distortion of the electronic density of states, induced by the resonant state, has been interpreted as the source of the high thermopower and figure of merit\cite{heremans-science} of Tl$_{0.02}$Pb$_{0.98}$Te at room temperature and above (e.g. at $T = 300$~K and hole concentration $p = 5-10\times 10^{19}$ cm$^{-3}$, $S$ is enhanced from $S \simeq 50 \mu$V/K, observed in Na-doped PbTe, to $S \simeq 125 \mu$V/K in Tl-doped PbTe\cite{ees-review}). Later on, the resonant enhancements of thermopower were observed in Tl doped Pb(Te-S) alloys\cite{ees-jaworski} or double (Si,Tl)-doped\cite{pbte-tl-si-nano} PbTe, with the peak $zT$'s of 1.6 at 700~K and 1.7 at 770~K, respectively.
We clearly see, that PbTe:Tl is an interesting system, still not fully understood, where intriguing physical properties are mixed with very good thermoelectric properties, which are of great practical interest.

The aim of this work is to study the modifications of the PbTe electronic bands and Fermi surface, induced by thallium doping. Although PbTe:Tl system has been widely studied, theoretical discussion of its band structure was limited mainly to the analysis of the density of states.\cite{ees-jaworski,mahanti-prl06,mahanti-prb08,ees-review,pbte-group3-calc} The only discussion of dispersion relations was done basing on supercell technique results,\cite{mahanti-prb08,pbte-group3-calc} which however, has several difficulties. At first, the original (rock salt in this case) crystal unit cell has to be replaced by the simple cubic one, changing the shape and size of the Brillouin zone (BZ), location of high symmetry points and leading to the band folding problem. Next, the cubic unit cell has to be multiplied to account for a doping level of the order of a few percent, which leads to the formation of hundreds of bands in the BZ, which are than difficult to analyze. Finally, due to the periodicity of the supercell, 
band structure of the simulated alloy always 
consist of perfect, well defined bands, with infinite lifetime of the Bloch states.
This can lead to a problem with the interpretation of the supercell results, particularly in the case of a system with resonant impurities, where electronic life time is limited by the strong impurity scattering.
In such a case sharp energy bands do not longer exist, as will be shown in the following section. Thus, the prediction of impurity bands formation, basing on supercells, should be verified using complementary methods.
Also, presence of the resonant impurity does not allow to discuss the band structure of PbTe:Tl basing on the rigid-band approximation applied to the pure PbTe, due to the strong modifications of the electronic structure.
To overcome these problems, we apply first principles calculations techniques developed for random alloys, i.e. the Korringa-Kohn-Rostoker method with coherent potential approximation\cite{kkr99,ebert-kkr2011} (KKR--CPA), where supercells are not used. For the first time Bloch spectral functions (BSFs), which replace dispersion relations in disordered system, are calculated for Tl$_{0.02}$Pb$_{0.98}$Te, and BSF intensity maps over the Brillouin zone (Fermi surface cross-sections) are presented. Results show that thallium indeed acts as a resonant impurity which e.g.  strongly modifies the density of states, however not creating an isolated impurity band close to the valence band (VB) edge. The modifications to the dispersion relations at first glance occur as strong blurring of electronic bands making the $E({\bf k})$ relations hard to define. 

\section{Computational details}

\begin{figure}[t]
\includegraphics[width=0.48\textwidth]{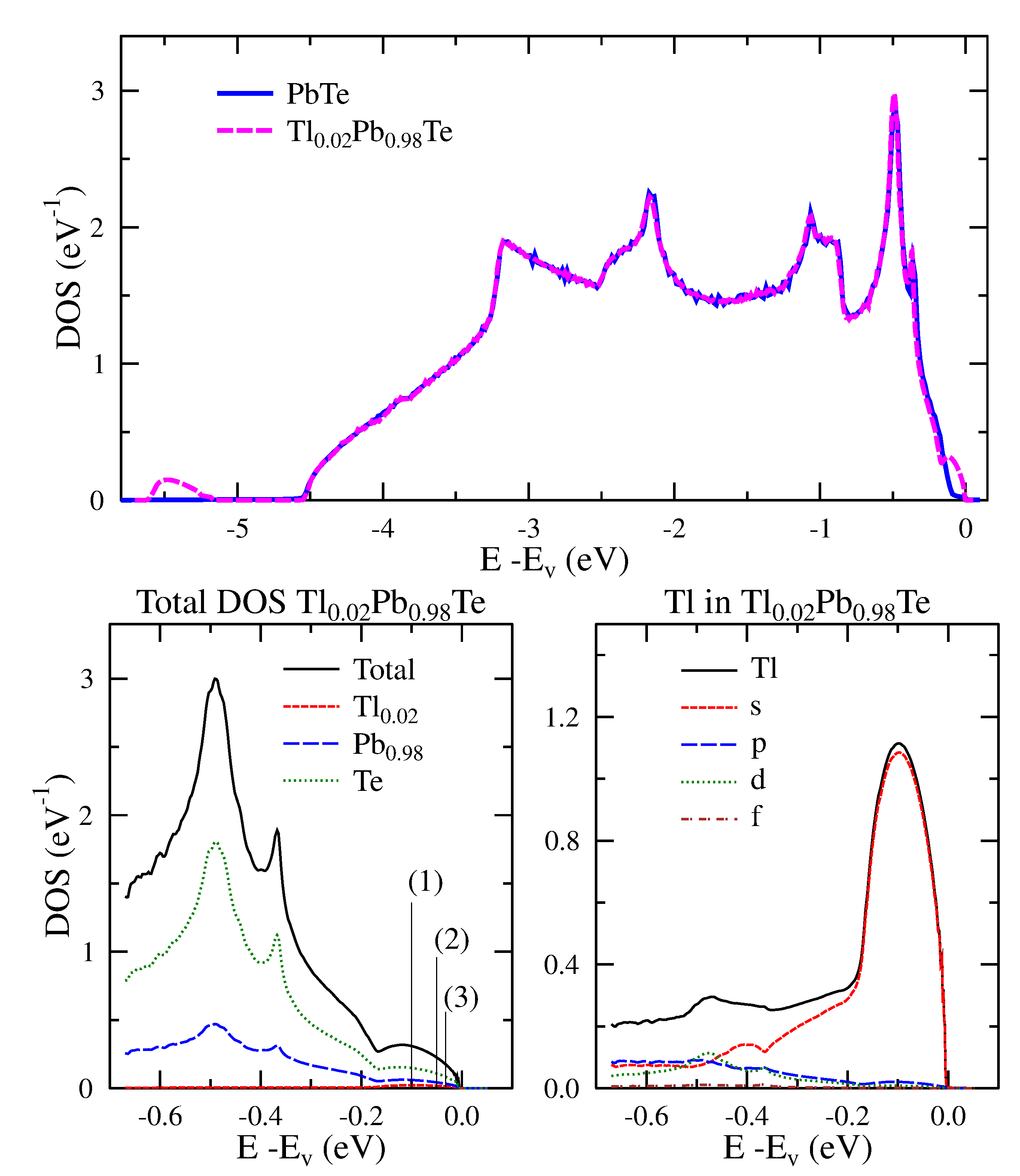}%
\caption{\label{dos}(Color online) Top panel: valence band DOS of PbTe and Tl$_{0.02}$Pb$_{0.98}$Te, with the two resonant peaks, around $-5.5$~eV and 0~eV, visible. $E_v$ is the energy of the valence band (VB) edge. Left bottom panel: total DOS close to the VB edge with atomic-decomposed contributions. Vertical lines mark the position of the chemical potential (at $T = 300$~K) for the carrier concentrations: (1) $3\times 10^{20}$ (nominal carrier concentration), (2) $1\times 10^{20}$, (3) $5\times 10^{19}$, all in cm$^{-3}$. Right bottom panel: partial DOS of thallium atom (per one Tl atom, i.e. not multiplied by Tl concentration).}
\end{figure}

Electronic structure calculations were performed using the full potential Korringa-Kohn-Rostoker (KKR) multiple scattering method, with the coherent potential approximation (CPA) used to simulate the chemical disorder. In present work, the Munich {\sc sprkkr} package\cite{sprkkr,ebert-kkr2011} was used, allowing us to include the spin-orbit interaction. The crystal potential was constructed in the framework of the local density approximation (LDA), using Vosko, Wilk and Nussair formula\cite{vosko} for the exchange-correlation part.
Experimental crystal structure and lattice parameters for the PbTe rock salt unit cell were used in computations:\cite{ravich-book} space group Fm-3m no. 225, a = 6.46~\AA. In order to increase the unit cell filling, empty spheres were added on the (8c) positions. High convergence limits were put on the self-consistent cycle ($10^{-5}$ Ry for the largest potential error), the angular momentum cut-off number $l_{max} = 3$ was used, regular k-mesh was used in calculations, using up to $62^3$ points for the self-consistent cycle and $192^3$ for the density of states (DOS) and Bloch spectral functions calculations. 
Lloyd formula\cite{ebert-kkr2011} was used to determine the Fermi level during the self-consistent calculations.
As far as the relativistic effects are concerned, spin-orbit interaction was included in all the calculations presented here. 

All the calculations were performed for the Tl$_{0.02}$Pb$_{0.98}$Te composition, which gives the nominal holes concentration $3 \times 10^{19}$~cm$^{-3}$. In different experimental studies, different carrier concentrations from the range $4 - 12 \times 10^{19}$~cm$^{-3}$ were reached for heavily Tl doped PbTe,\cite{heremans-science,ees-jaworski,matsushita-supercond} being always lower than the nominal ones. This discrepancy was discussed in literature usually in terms of the so called self-compensation model,~\cite{ravich-review, ravich-review2} also charge Kondo  model was successfully applied to account for it.~\cite{zlatic-prl2012} In present paper, to show the spectral functions corresponding to the experimentally observed carrier concentrations ($5 \times 10^{19}$ and $1 \times 10^{20}$~cm$^{-3}$ were selected) Fermi level was rigidly shifted.

\section{Results and discussion}\label{sec:bsf}

\begin{figure}[b]
\includegraphics[width=0.50\textwidth]{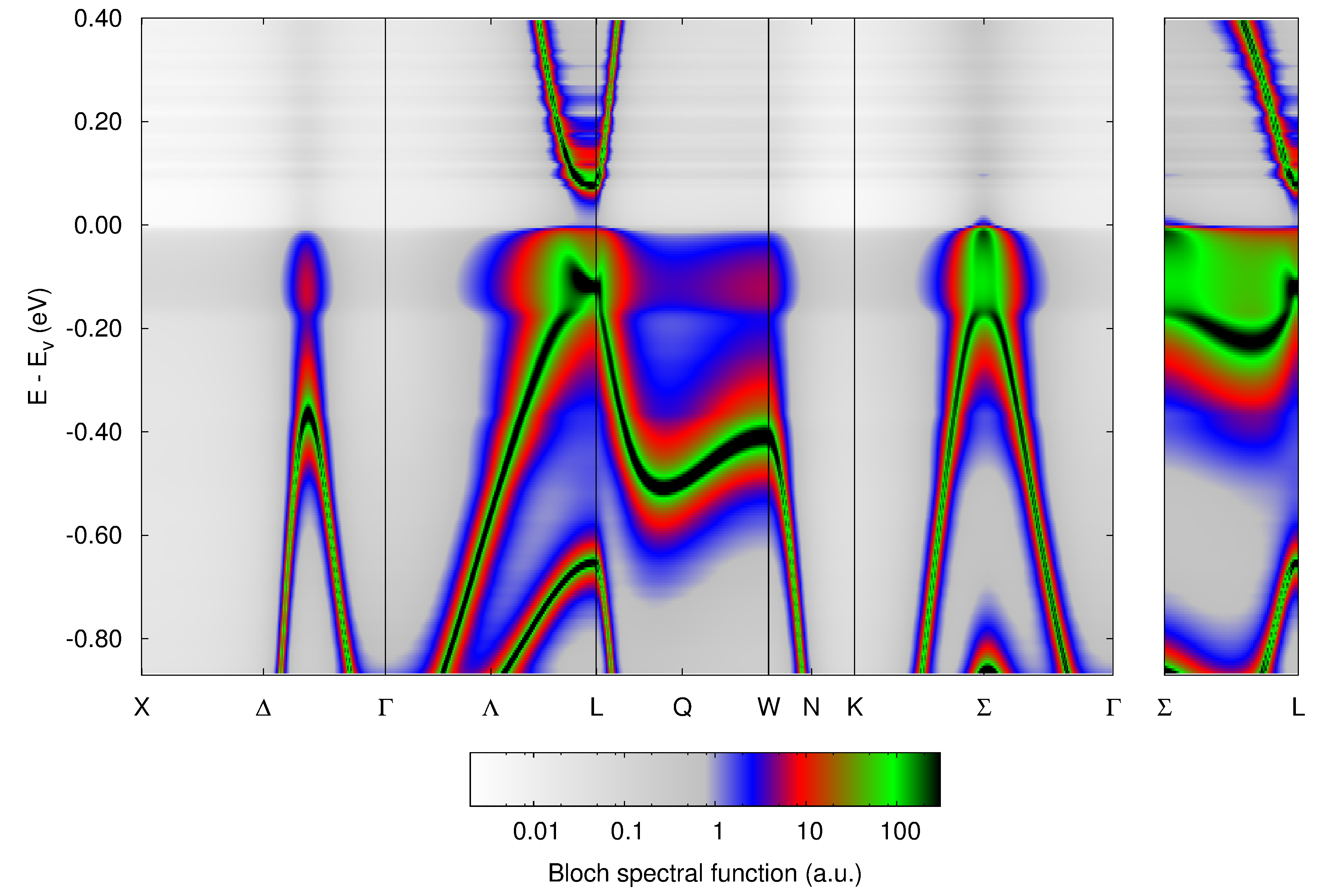}%
\caption{\label{bloch}(Color online) The two-dimensional projections of Bloch spectral functions of Tl$_{0.02}$Pb$_{0.98}$Te in the high symmetry directions. Here, and in next figures, the black color corresponds to BSF value greater than 300 a.u., i.e. 22 eV$^{-1}$. The right panel, with the $\Sigma - L$ direction, has the same energy scale as the left one.}
\end{figure}

Figure~\ref{dos} shows the density of states (DOS) of the valence band of Tl$_{0.02}$Pb$_{0.98}$Te. The bottom panels show the total and thallium DOS close to the valence band edge. In the top panel, where DOS of pure PbTe is also plotted, we see the formation of the resonant states, triggered by the Tl 6s electrons, as the peaks in DOS close to -5.5 eV and the valence band edge, in agreement with previous studies.\cite{ees-jaworski,ees-review,mahanti-prl06,takagiwa-tlpbte}.

\begin{figure*}[t]
\includegraphics[width=0.95\textwidth]{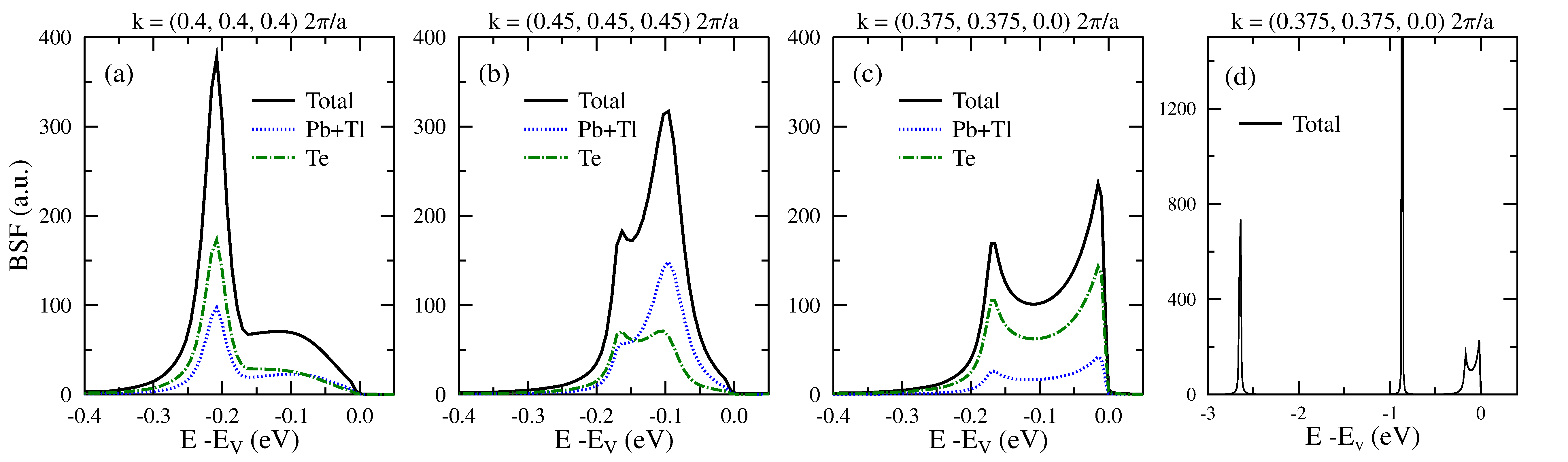}%
\caption{\label{bloch-k}(Color online) Bloch spectral functions of Tl$_{0.02}$Pb$_{0.98}$Te plotted at {\bf k} = $(0.4, 0.4, 0.4)$, $(0.45, 0.45, 0.45)$ (close to $L$ point), and $(0.375, 0.375, 0.0)$ ($\Sigma$ point), all in units 2$\pi$/a. Contributions to BSF from Tl+Pb and Te sites are plotted in colors.}
\end{figure*}

Three valence electrons of Tl, two 6s and one 6p are then occupying: (first 6s) the ''hyper--deep'' state, located about 5.5 eV below the valence band (VB) edge, (6p) the main valence band (from -4.5 eV to VB edge) and (second 6s) the ''deep'' resonant state, at the VB edge (the ''hyper--deep'' and ''deep'' names are taken after Ref.~\onlinecite{mahanti-prb08}). This 6s state splitting is another unique feature of the group-III impurities in PbTe,\cite{mahanti-prl06} since typically one would expect the 6s state to be degenerate. Presented picture rather supports  thallium being in the Tl$^+$ state, since Tl$^{3+}$ state would require excitation of the electron from  the low-lying hyper-deep state.\cite{mahanti-prl06}
In the following part we will focus mainly on the energy region close to the valence band edge.
The vertical lines in the bottom left Fig.~\ref{dos} mark the position of the chemical potential (at $T = 300$~K)
for the carrier concentrations: (1) $3\times 10^{20}$~cm$^{-3}$ (nominal carrier concentration, i.e. assuming one hole per Tl atom $\simeq$ theoretical Fermi level), (2) $1\times 10^{20}$cm$^{-3}$, (3) $5\times 10^{19}$cm$^{-3}$, all scanning the resonant hump.

Presented results are in good agreement with our previous, semi-relativistic calculations\cite{ees-jaworski,ees-review} which were done using independent implementation of the KKR-CPA method.\cite{kkr99,bw-fe2p,bw-y4co3}
The main relativistic effect is the reduction of the energy separation between the valence and impurity or conduction states. This manifests as (i) the reduction of the {\it neck} between the resonant hump and main valence band, and (ii) the reduction of the band gap, which for Tl$_{0.02}$Pb$_{0.98}$Te is about 0.6 eV in semi-relativistic calculations\cite{ees-jaworski} and 0.1 eV when spin-orbit interaction is included (see also Fig.~\ref{bloch}). Note, that we have verified that gradient corrections, as incorporated in the GGA Perdew-Burke-Ernzerhof functional\cite{pbe} do not change the gap value. 
Thus, the error in the band gap magnitude for PbTe:Tl is similar to the undoped PbTe case, which was discussed in literature in the past\cite{pbte-papa,pbte-zunger} (experimental low temperature value of the band gap in PbTe is $E_g \simeq 0.2$ eV).

Let us recall, that the DOS hump is formed due to the presence of Tl 6s resonant DOS peak, however, the strong hybridization of the Tl 6s states with Pb and Te p-states makes the Pb and Te contributions to the total DOS in the peak dominating. This was earlier tentatively ascribed as the important feature for the good thermoelectric properties of this system,\cite{ees-jaworski,ees-review} since it showed the balance between localization of the impurity states and delocalization of the Bloch states of the host crystal. Too localized impurity states, which would give the steep DOS favoring high thermopower on one hand, would reduce the electrical conductivity and impurity contribution to the thermopower (see also discussion of the impurity band formation below). Results, presented in the following part of this paper, will give complementary arguments for the delocalization of the resonant impurity states.

To be able to discuss the PbTe:Tl band structure, we have calculated the Bloch spectral density functions $A^B(\mathbf{k},E)$ (BSFs), which replace the $E({\bf k})$ dispersion relations in the disordered solids. BSFs
are related to the Fourier transformed  configurationally averaged electron's Green's function:\cite{ebert-bloch,faulkner-cpa,ebert-kkr2011}
\begin{eqnarray}
A^B(\mathbf{k},E) &=& -\frac{1}{\pi N} \sum_{i,j}^N e^{i \mathbf{k}(\mathbf{R}_i - \mathbf{R}_j)} \times \nonumber \\&& {\rm Im \ Tr} \int_{\Omega} d^3r \langle G (\mathbf{r} + \mathbf{R}_i, \mathbf{r} + \mathbf{R}_j)\rangle \label{eq:bsf}
\end{eqnarray}
where $i,j$ are lattice site indices, and $\Omega$ the unit cell volume. Since the imaginary part of the Green's function is related to the density of states, it can be shown that $A^B(\mathbf{k},E)$ has an interpretation of a
{\bf k}-resolved DOS:\cite{ebert-bloch, faulkner-cpa}
\begin{equation}
DOS(E) = \frac{1}{\Omega_{BZ}} \int_{BZ} d^3k A^B(\mathbf{k},E).
\end{equation}
For perfect crystal, i.e. an ordered medium, BSF is a Dirac delta function, being equal to one only at the ({\bf k}, E) points where electron in band $\nu$ has energy eigenvalue $E_{\nu,\mathbf{k}}$, and here BSF define the usual dispersion relation:
\begin{equation}
A^B(\mathbf{k},E) = \sum_{\nu} \delta(E - E_{\nu,\mathbf{k}}).
\end{equation}

If the influence of the impurities on energy bands is relatively weak or moderate, BSF takes the form of Lorentz function, with the full width at half maximum (FWHM) value $\Gamma$ corresponding to the life time of the given electronic state:\cite{gyorffy-cu_ni}
\begin{equation}\label{eq:time1}
\tau = \frac{\hbar}{\Gamma}
\end{equation}
In such a case we can still define {\it virtual} energy band in alloy, with the band center at the energy, where BSF lorentzian has maximum, and a bandwidth corresponding to $\Gamma$. In this case the effect of alloying leads to widening of bands. This description is equivalent to the complex energy band technique\cite{butler-kubo,kamil-prb}, where electronic energy eigenvalue has real part (the band center) and imaginary part, corresponding to the life time:\cite{butler-kubo}
\begin{equation}\label{eq:time2}
\tau = \frac{\hbar}{2{\rm Im}(E)}.
\end{equation}
Comparing Eq.~\ref{eq:time1} to Eq.~\ref{eq:time2} we see, that $\Gamma = 2{\rm Im}(E)$.
The stronger is the electron scattering by impurities, the wider is the BSF, leading to strong widening of energy bands and decreasing the electronic life-time. That is especially the case of resonant scattering and resonant impurities, where BSF can loose the sharp lorentzian shape, making the energy band center hard to define. The classical examples of such cases are the transition-metal noble-metal alloys, e.g. Cu-Ni.\cite{gyorffy-cu_ni}

\begin{figure*}[t]
\includegraphics[width=0.95\textwidth]{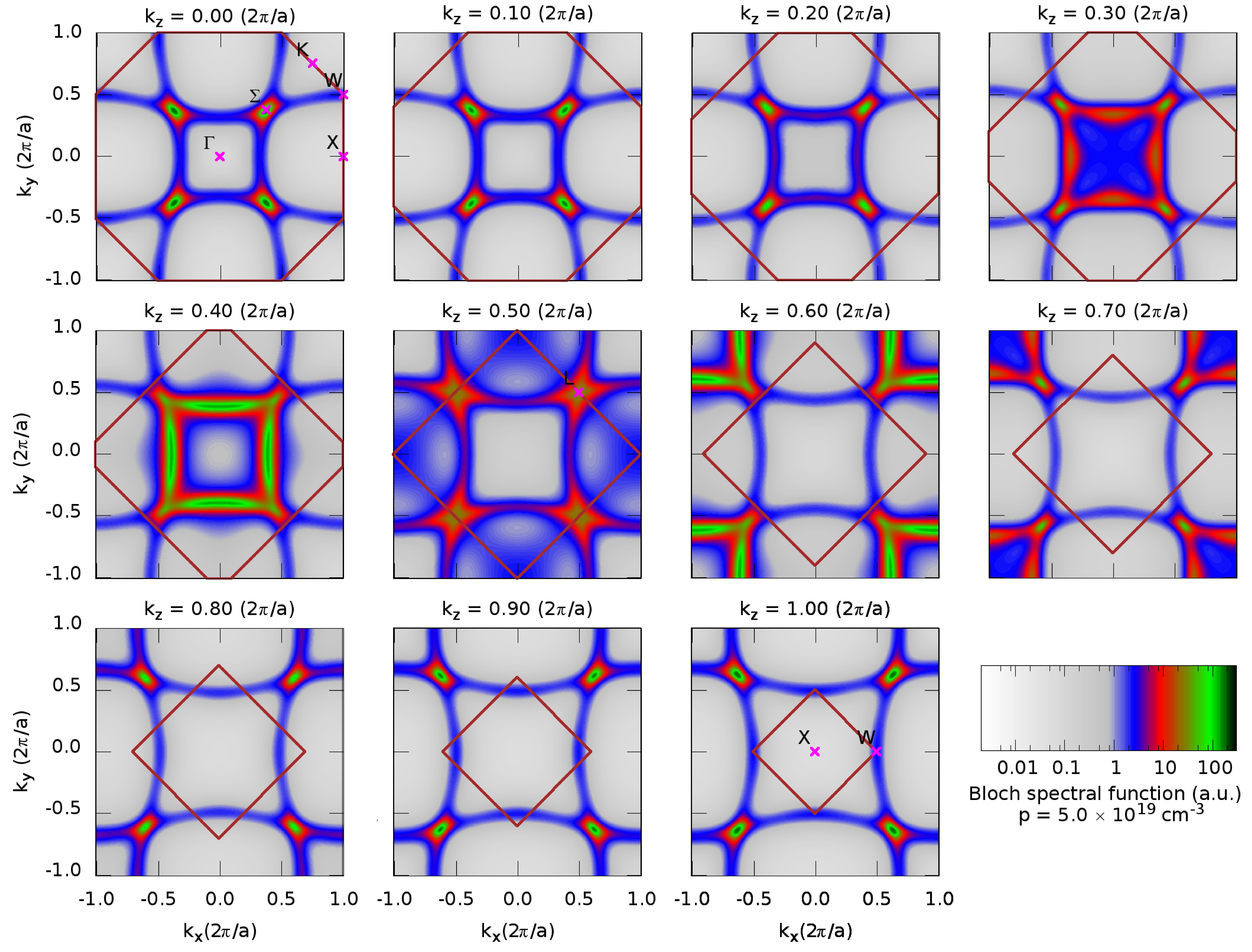}%
\caption{\label{slice1}(Color online) Cuts through the BZ with Bloch spectral function values marked by colors, plotted in logarithmic scale (Fermi surface cross-sections) for the carrier concentration $p = 5\times 10^{19}$ cm$^{-3}$.}
\end{figure*}

\begin{figure*}[t]
\includegraphics[width=0.95\textwidth]{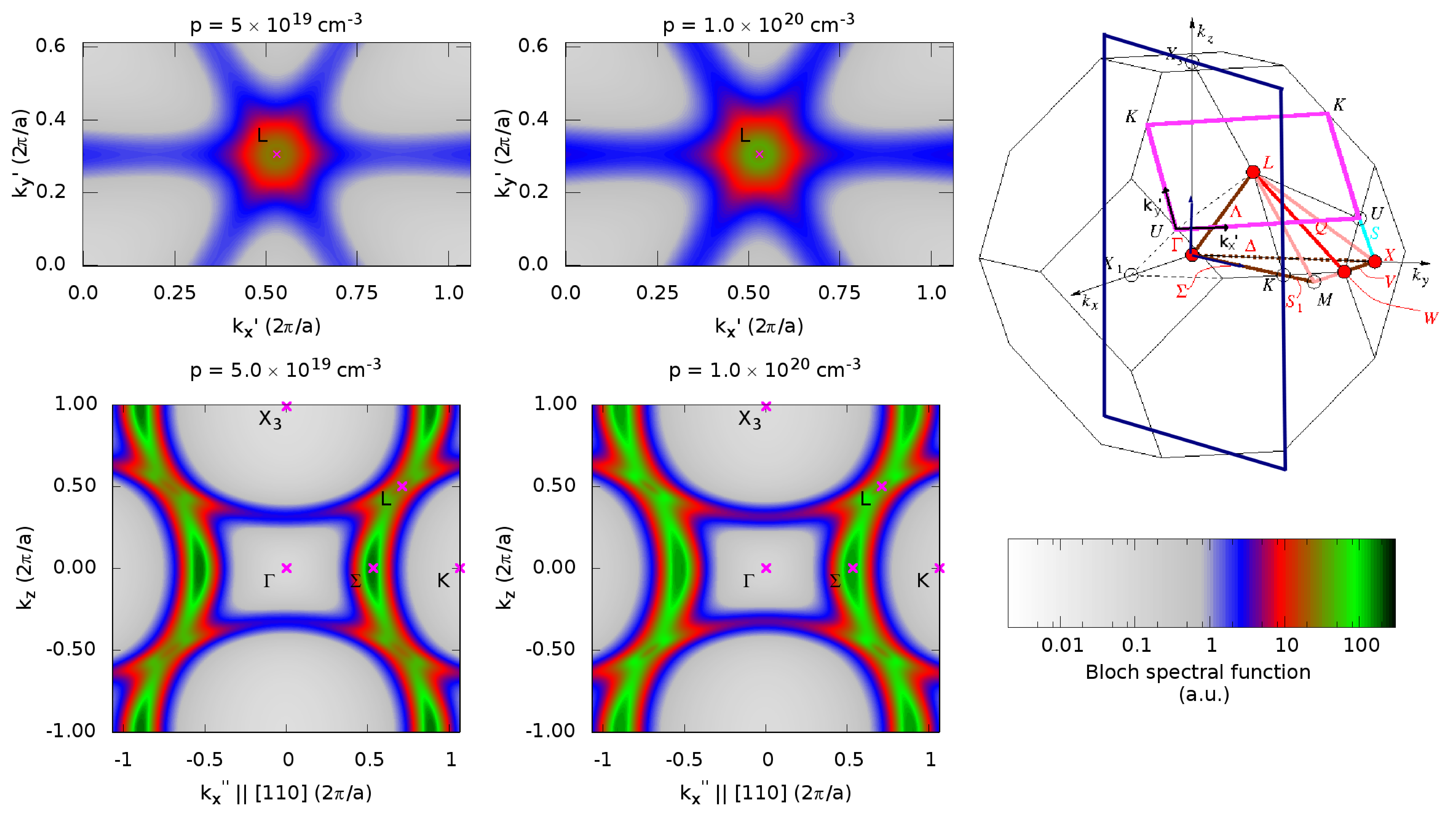}%
\caption{\label{slice2}(Color online) Cuts through the BZ with Bloch spectral function values marked by colors, plotted in logarithmic scale (Fermi surface cross-sections) for the carrier concentrations $p = 5\times 10^{19}$ and $1.0\times 10^{20}$cm$^{-3}$. Top panels show the rectangle between $K-K-U-U$ points in BZ, marked in magenta in the BZ picture, bottom panels show the rectangle lying between $X_3$ and $K$ points, marked in navy in the BZ picture.}
\end{figure*}

The two-dimensional projections of BSFs, in the $(k, E)$ variables along the high symmetry directions (in analogy to the usual dispersion relations) are presented in Fig.~\ref{bloch}, with the $A^B(k,E)$ values marked by colors. We see, that for $E < -0.3$~eV (zero is at the valence band edge), i.e. below the resonant hump in DOS, {\it virtual} energy bands could be easy identified, as the points where BSFs have maxima. However, between -0.3 eV and valence band edge resonant state develops, mainly around $L$ point, but also states from the so-called heavy-hole band at $\Sigma$ point are coupled. Strong distortions of the electronic bands are clearly visible, e.g. the top of the valence band is moved from $L$ point, and electronic states forms ''clouds'' around $L$ and $\Sigma$.
If we do not limit the discussion to the high-symmetry directions, important feature of PbTe:Tl bandstructure can be noticed. The $L$ and $\Sigma$ points are directly connected by the valence band, and the strongest effect of Tl doping is observed for this particular band, which is additionally plotted in Fig.~\ref{bloch}. Note, that $L-\Sigma$ connection can also be seen as a bridge between different $L$-points, building a network of $L-\Sigma-L$ connections at the doping levels we are interested in ($p > 4\times 10^{19}$~cm$^{-3}$), see also Fig.~\ref{slice1} and Fig.~\ref{slice2}.
This $L - \Sigma - L$ connection is also present in undoped PbTe, although this feature is rarely discussed in literature. Recently this connection was identified to exist in all lead chalcogenides and described in terms of reduced dimensionality,\cite{singh-2d} which was proposed to be responsible for the good thermoelectric properties of these materials. Also, temperature dependence of the band structure of the p-type PbTe-PbSe alloy has been attributed\cite{heremans-topol} to the topological modifications of this connection, leading to the formation of ''topological electrons''. Here we see that Tl doping strongly affects this band leading to its enormous blurring, so sharp $E({\bf k})$ relation becomes hard to define. 

The BSFs at two points around $L$ and at $\Sigma$ points are plotted in Fig.~\ref{bloch-k}. Fig.~\ref{bloch-k}(d) shows BSFs at $\Sigma$ in larger scale, the deeper lying peaks come from the rather well-defined virtual bands, located around 2.5 and 1 eV below the valence band edge. Here the 
BSFs are sharp and lorentzian-like, which is 
definitely neither 
(a) nor (c) case. The non-regular shape of the spectral function makes it ambiguous to define the electron life time, associated with such blurred electronic states. Nevertheless, taking the FWHM to be about $\Gamma = 0.1 - 0.2$~eV we get $\tau \simeq 3.3 - 6.6 \times 10^{-15}$~s, thus we should be able to safely put $\tau$ in the range $\tau \simeq 10^{-15} - 10^{-14}$~s. Interestingly, at $\Sigma$ point we may notice higher contribution to BSF from Te atoms, which is another proof of the strong mixing of the introduced impurity states with Te states, underlying this unique resonant impurity character.

The unusual character of the PbTe:Tl band structure, with such 'blurred' and non-lorentzian spectral functions, raise the question on the interpretation of the Hall effect measurements, often used to measure the carrier concentration $n$ with the free-electron formula $R_H = -1/(n e)$.
As was mentioned before, in this system the Hall carrier concentration is always lower, than the nominal one. Among the potential explanations of this fact, a possibility that it is due the RL-induced strong modifications of the Fermi surface, has never been taken into account, and should be addressed in the future studies.

To better visualize the redistribution of electronic states in the Brillouin zone, a series of constant-energy-surface plots -- cross-sections through the alloy's ''Fermi surface'' (in quotes, since it is not longer a surface) -- have been calculated. The energy, corresponding to the concentration of holes $p = 5\times 10^{19}$ cm$^{-3}$ has been selected, as it corresponds to that measured experimentally in the Tl$_{0.02}$Pb$_{0.98}$Te samples.\cite{heremans-science} 

Fig.~\ref{slice1} shows the set of $k_x-k_y$ BSFs intensity plots on a set of planes perpendicular to $k_z$ axis, starting from $k_z = 0$ up to the BZ top. In each of the figures, the BZ boundaries are marked, i.e. the points inside belong to the first BZ. High symmetry points are also marked, see the BZ picture in Fig.~\ref{slice2}. Here we see how the high values of BSFs form ellipsoids around $L$ and $\Sigma$ points, and how the connections between these points are developed, with the highest intensity seen for $k_z = 0.4$ (2$\pi$/a).

Two different sets of BSFs projections are presented in Fig.~\ref{slice2}, this time the energy corresponds to two carrier concentrations, $p = 5\times 10^{19}$ and $1.0\times 10^{20}$cm$^{-3}$ (higher concentrations were obtained e.g. in Tl-doped PbTe$_{1-y}$S$_y$ samples (Ref.~\onlinecite{ees-jaworski})  and in Tl$_{x}$Pb$_{1-x}$Te single crystals in Refs.~\onlinecite{matsushita-supercond,matsushita_res_kondo}). Top panels show the rectangle section of the BZ boundary hexagon, with $L$ point lying in the center, marked also on BZ picture on the right. We clearly see the almost circular base of the L-point ellipsoids, with all the inside surface filled with electronic states. Bottom panels show the Fermi surface cut with the X$_3-\Gamma-$K plane, where the $\Sigma$ ellipsoids and their connection to the L points are well visible. Also here the ellipsoids are filled with states inside. 
The smearing of the alloy Fermi surface makes the $L - \Sigma - L$ connections so wide, that their intensity (although small) are even visible for $k_z = 0$ (Fig.~\ref{slice1}), thus they are forming a face of a cube, since have a noticeable intensity for all $k_z$ up to the BZ top. 
Roughly we may conclude, that in Tl-doped PbTe the alloy ''Fermi surface'' skeleton is based on the Fermi surface of PbTe, however strongly modified due to the presence of the resonant impurity. If Tl was rigid-band-like impurity, the Fermi surface would consist of empty tubes, building the $L-\Sigma-L$ connections, like those presented for PbTe in Refs.~\onlinecite{singh-2d,singh-pbte}. Here, due to the resonant band distortion (Fig.~\ref{bloch}) the tubes are filled with additional electronic states, which is the reason for the occurrence of the excess density of states close to the VB edge, as seen in Fig.~\ref{dos}. 
Originally~\cite{heremans-science} the effect of thermopower enhancement by the RL was qualitatively explained in terms of the local increase of DOS due to the RL-induced DOS distortion, since approximately for $p$-type semiconductor $S \propto DOS(E_F)/\int_{E_F}^{E_V} DOS(E)$, i.e. the higher is the DOS at given number of carriers, the higher is the thermopower  
(see, also Ref.~\onlinecite{ees-review}). Analysis of the spectral functions presented here can give us another look on this feature -- in PbTe:Tl RL creates a redistribution of the electronic states, increasing the number of states around $L$ point and in the $L-\Sigma-L$ band. Thus, we have additional states around electronic bands which have good thermoelectric properties. This can improve the thermoelectric performance of the system roughly in a similar way as increasing the band degeneracy, like the band convergence effect, widely discussed\cite{kamil-prb,prl-mg2si-uher,fedorov-crc} for Mg$_2$(Si-Sn) or Pb(Te-Se).\cite{pei-nature} However, this is only qualitative comparison, since the resonant mechanism is of course much different: instead of moving the existing bands, the interaction between the introduced impurity electronic states and the host crystal states transfers electronic states from the deeper energies towards the VB edge, creating a ''cloud'' of states around the original valence band. 

In this view, the positive role of thallium in improving the thermoelectric properties may come both, from the resonant character of the impurity {\it and} the specific band structure of PbTe. RL itself probably would not lead to such a good TE properties, and vice versa, in spite of PbTe unique band structure, without RL one cannot explain the enhancement in thermopower of PbTe:Tl, against e.g. Na:PbTe.~\cite{ees-jaworski} 
The important effect is, that according to our results, Tl in PbTe does not form a new, isolated impurity band close to the VB edge -- spectral functions do not show any signs of new impurity band formation in Fig.~\ref{bloch}.
This is in contrast to the previous calculations, where formation of Tl-like impurity band was detected, as a natural consequence of the supercell technique.\cite{mahanti-prb08,pbte-group3-calc} 
But as was mentioned in the Introduction, due to multiplication of the unit cell and enforced periodic conditions, supercell technique has to predict the formation of many new, sharp energy bands.
The comparison between KKR-CPA spectral functions and supercell Bloch energy bands would require from the latter at least probing several of the possible configurations of the host and impurity atoms to calculate the configurational average (which is done in CPA). After that, supercells would possibly yield similar band widening effect, as we got from CPA, but still the problem of unfolding the BZ would remain (especially difficult for the single, isolated band). Since such calculations have not been done so far, we have to state that both techniques give different prediction for the impurity band formation, in spite of giving similar DOS picture.

\begin{figure}[htb]
\includegraphics[width=0.49\textwidth]{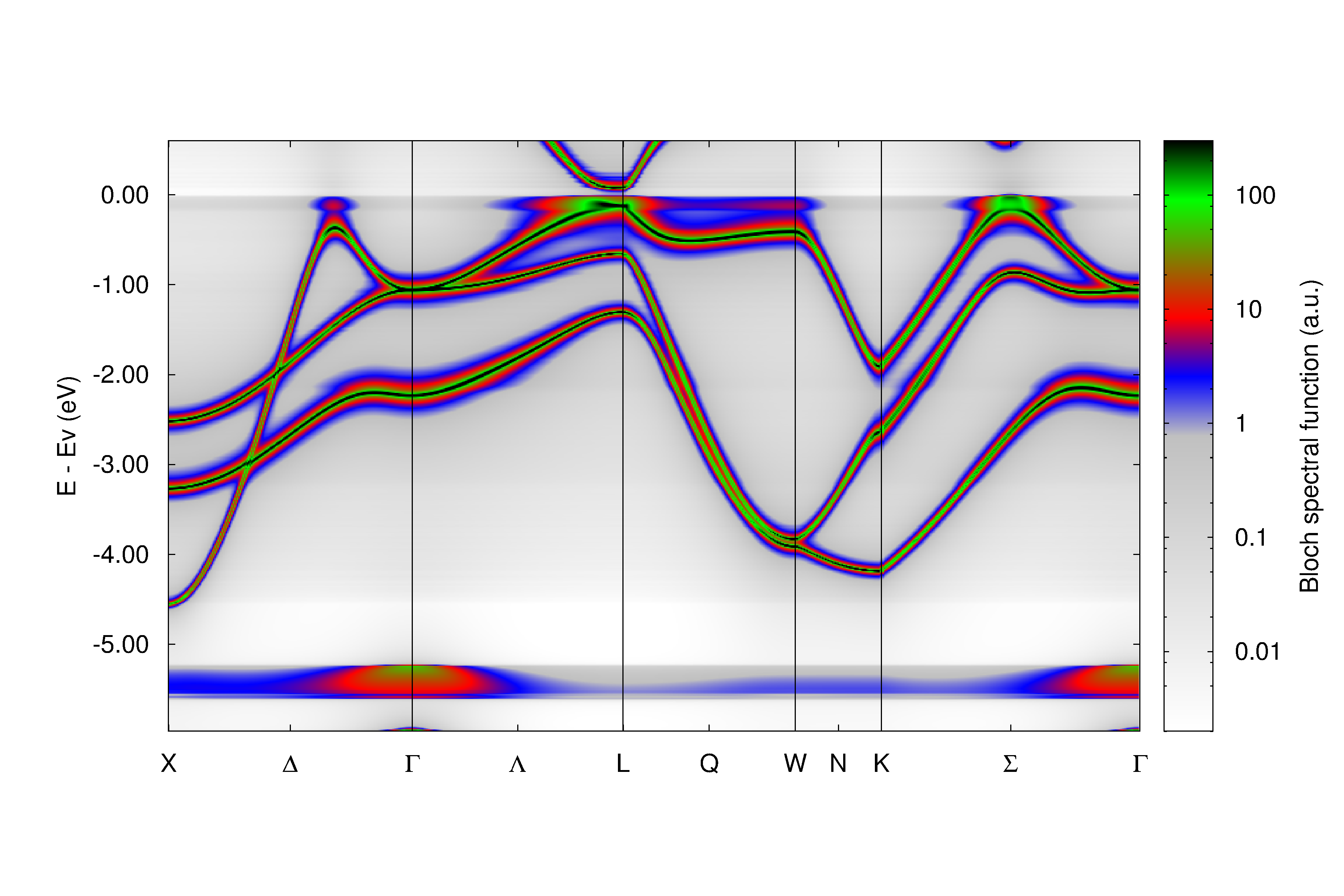}%
\caption{\label{bloch_all}(Color online) Bloch spectral functions of Tl$_{0.02}$Pb$_{0.98}$Te plotted in a broader energy range. The BSF of ''hyper-deep'' state around -5.5 eV is visible, and resembles the isolated impurity band, in contrast to the RL located close to the VB edge.}
\end{figure}

Situation is different for the hyper-deep state, lying around -5.5 eV. Here, Bloch spectral function resembles a narrow isolated impurity band, which can be seen in Fig.~\ref{bloch_all}, where BSFs are plotted in a broad energy range. 
The flat character of its spectral function also confirms the more-localized character of the hyper-deep state, comparing to the deep-defect state. These results are in agreement with the supercell bandstructure and charge density analysis.\cite{mahanti-prb08}. 

The lack of the isolated impurity band close to VB is especially important in the context of TE properties. The explanation of the thermopower enhancement in PbTe:Tl, if the impurity band was formed, would be difficult. In such a case, the resonant peak in DOS would require the impurity band to be rather narrow. Such a band would likely have small conductivity, thus its contribution to the total thermopower would be small, since approximately for the two-band system of the host and impurity bands, the thermopower is a weighted average\cite{ees-review,barnard} $S = (\sigma_{\rm imp} S_{\rm imp} + \sigma_{\rm host} S_{\rm host})/(\sigma_{\rm imp} + \sigma_{\rm host})$. Thus, when $\sigma_{\rm imp} \ll \sigma_{\rm host}, S \simeq S_{\rm host}$. Recently, Goldsmid has shown,\cite{goldsmid-impurity} that special conditions are required for the impurity band to improve the TE properties. Now as we have shown that there is no isolated impurity band in PbTe:Tl, and that impurity states are not localized, the 
enhancement of thermopower in this system should be easier to understand. 
However, the direct proof of the positive role of the resonant state on TE properties should come from the first-principles calculations. Such computations are now being performed, with the help of Kubo-Greenwood formalism,\cite{kubo57,greenwood58,ebert-kkr2011} since for the system without well-defined energy dispersion relations $E({\bf k})$ one cannot 
use the Boltzmann approach. Our preliminary results\cite{ict2013} show, indeed, the high thermopower of Tl$_{0.02}$Pb$_{0.98}$Te system, and this will be the subject of the forthcoming paper.

\section{Summary}

To summarize, in present work the results of KKR-CPA calculations of the Bloch spectral functions for Tl$_{0.02}$Pb$_{0.98}$Te have been presented. The resonant state, induced by thallium atoms in lead telluride, strongly affects the electronic bands close to the valence band edge, leading to their substantial blurring and widening. Resonant state develops up to about 0.3 eV below the valence band edge, and in this energy range, especially in the $\Sigma - L$ direction in the Brillouin zone, well defined electronic bands do not exist. Impurity states, introduced to PbTe with Tl atoms, strongly interact with the host states of the crystal, creating ''clouds'' of states around the initial valence band. As a consequence, the alloy ''Fermi surface'' consists of tubes connecting $L - \Sigma - L$ points, however, instead of being empty (as in PbTe case), they are filled inside with additional electronic states, being of mixed impurity and host character. This effect is reflected in the DOS as an increase of the 
number of states close to VB edge. In view of these results, the effect of the thermopower enhancement, observed 
experimentally in 
PbTe:Tl, can be roughly compared with the effect of the increase in band degeneracy, known to improve thermoelectric properties. Also, in view of present results, formation of additional, isolated impurity band close to the VB edge is unlikely. 

\begin{acknowledgments}
This work was partly supported by the Polish National Science Center (project no. DEC-2011/02/A/ST3/00124).
I thank Dr. Sven Bornemann and Prof. Hubert Ebert for the discussion of the full-potential mode of the {\tt SPRKKR} package.
\end{acknowledgments}

\bibliography{ref-pbte_bloch}

\end{document}